\documentclass[aps,pra,showpacs,twocolumn]{revtex4}
\usepackage{epsfig}
\begin{document}
\title{Optimization of security costs in nested purification protocol}
\author{Radim Filip\thanks{e-mail filip@thunder.upol.cz}}
\affiliation{
Department of Optics,
Palack\'y University,
17. listopadu 50, 772 00 Olomouc,
Czech Republic}
\date{\today}

\begin{abstract}
We analyse security costs in one segment 
of nested purification protocol in a large quantum cryptography network,
employing the quantum switchers and repeaters.
We demonstrate that exponential or even super-exponential grow of
entanglement resources occurs in dependence on
number of the network switchers. For this reason,
an optimization in the nested strategy is suggested, preventing
a stronger than the exponential grow of entanglement resources.
\end{abstract}

\pacs{PACS number(s):00}

\maketitle

Recently, quantum cryptography link using entangled states 
\cite{Ekert91,Bennett92} has been experimentally realized \cite{Naik00},
which is basic step to futured quantum cryptographic network.
To construct this network, original Ekert's
protocol has been extended by the quantum memories \cite{Biham96}
and quantum switchers \cite{swapp}.
Quantum memories carefully store the entangled states
before key distribution and the
quantum switchers are able to swap
entanglement and consequently, establish a secret communication
between distant users.
However, the entangled states transmitted between the users or
stored in the quantum memories
cannot be precisely protected against undesirable decoherence 
or the deliberate eavesdropper attacks.
Due to these influences, the entanglement {\em exponentially}
vanishes with increasing distance between the users.
Fortunately, quantum cryptography based on Ekert's protocol 
presents a potential advantage,
since the users could regenerate the secure key by
implementing a purification protocol \cite{Bennett,Deutsch}. 
Quantum purification is able to 
extract sufficiently entangled pair of photons from a number of
the pairs with weak entanglement, utilizing
only the local operations and classical communications.
Thus, the links between users
in the networks achieve a sufficient security and
the almost maximally entangled pairs can be 
stored in the quantum memories for the next usages.

Practical implementation of the swapping/purification idea based on
the quantum repeaters in connection points of network extended standard 
purification to 
{\em nested} purification protocol (NPP) \cite{Briegel98,Dur99}. 
We can assume a link consisting of total number $N-1$ of the switchers
connecting $N$ pairs of imperfectly entangled states, having non-unit
fidelity with maximally entangled state.
On the first level, we implement the quantum
repeaters in the check points with $L$
switchers between two repeaters. To purify one pair with required
amount of entanglement between nearest 
repeaters, 
we need a certain number $M$ of copies that we construct in parallel
fashion, as is depicted in Fig.~1.
The total number of elementary pairs is $M(L+1)$. On the second level,
we repeat this procedure considering the outcomes of first level as
input pairs for second level.
In summary, the total number of elementary pairs will be \cite{Briegel98}
\begin{equation}
R=N^{\log_{L+1}M+1}
\end{equation}
which shows that the resources grow {\em polynomially} with the distance $N$.
However, the number of needed entangled pairs $M$ per segment and per
one nesting level depends on number of switchers $L$, on
fidelity of pairs between the nearest users
and required fidelity of outgoing pairs between the distant users.
To simplify discussion, we require the same value of a long-distance
fidelity as the nearest users fidelity and denote it
as ``working'' fidelity. It has been previously shown
\cite{Briegel98,Dur99} that for $L=2$, the optimal ``working'' fidelity 
$F\approx 0.95$ requires minimal number of four pairs
per segment and nesting level.
However, if we want to create entanglement over arbitrary distance
(large $L$), there is still a question,
how much weakly entangled pairs
are needed between two repeaters for purification up to working fidelity.
Shortly, what is the relation $M=M(L,F)$ in the nested purification protocol?

In this paper, we present an analysis of these security costs,
assuming an almost perfect entanglement between the users
in the neighbourhood. To distribute entanglement between the
distant users in cryptographic network,
an idealized swapping/purification protocol will be assumed.
These procedures can be implemented probabilistically, with the help of
only linear optics elements \cite{linear}.
We demonstrate a new and important fact,
that the number of imperfectly entangled states between two
quantum repeaters grows {\em exponentially} or even {\em super-exponentially}
with the number of implemented switchers.
It is a cost of {\em exponential} decoherence elimination,
which must be paid to achieve a secure key distribution.
For this reason, we suggest an {\em optimized} nested 
purification protocol (ONPP),
which prevents the super-exponential grow of the security costs.
It will be important for a construction of the
quantum cryptographic network, expected in the next future.


Now we will theoretically 
discuss both the entanglement swapping and distillation procedures in one
segment of nested purification protocol, as is depicted in Fig.~1.
We assume that the nearest users $(i,j=i+1)$ share two-qubit state
\begin{equation}\label{genst}
\rho^{(i,j)}=\sum_{k,l}B^{(i,j)}_{k,l}|B_{k}\rangle\langle B_{l}|
\end{equation}
which is given in the Bell basis 
\begin{eqnarray}\label{Bellst}
|B_{1,2}\rangle&=&\frac{1}{\sqrt{2}}(|00\rangle\pm|11\rangle),\nonumber\\
|B_{3,4}\rangle&=&\frac{1}{\sqrt{2}}(|01\rangle\pm|10\rangle).
\end{eqnarray}
Note that diagonal elements $B_{k,k}=\langle B_{k}|\rho|B_{k}\rangle$ 
are the fidelities $B_{k}$ with particular Bell states (\ref{Bellst}).
First, we discuss a propagation of entanglement in the network
utilizing the standard entanglement swapping \cite{swapp} 
between two links 
$(i,i+1)$ and $(i+1,i+2)$.
Assuming Bell state analyser, 
which, at least conditionally, distinguishes between four Bell states
(\ref{Bellst}), the diagonal elements $B_{k,k}=B_{k}$
change along rule
\begin{widetext}
\begin{eqnarray}\label{swapp}
B^{(i,i+2)}_{1}&=&B^{(i,i+1)}_{1}B^{(i+1,i+2)}_{1}+
B^{(i,i+1)}_{2}B^{(i+1,i+2)}_{2}+
B^{(i,i+1)}_{3}B^{(i+2,i+3)}_{3}+B^{(i,i+1)}_{4}B^{(i+1,i+2)}_{4},
\nonumber\\
B^{(i,i+2)}_{2}&=&B^{(i,i+1)}_{1}B^{(i+1,i+2)}_{2}+
B^{(i,i+1)}_{2}B^{(i+1,i+2)}_{1}+
B^{(i,i+1)}_{3}B^{(i+1,i+2)}_{4}+B^{(i,i+1)}_{4}B^{(i+1,i+2)}_{3},
\nonumber\\
B^{(i,i+2)}_{1}&=&B^{(i,i+1)}_{1}B^{(i+1,i+2)}_{3}+
B^{(i,i+1)}_{3}B^{(i+1,i+2)}_{1}+
B^{(i,i+1)}_{2}B^{(i+1,i+2)}_{4}+B^{(i,i+1)}_{4}B^{(i+1,i+2)}_{2},
\nonumber\\
B^{(i,i+2)}_{1}&=&B^{(i,i+1)}_{1}B^{(i+1,i+2)}_{4}+
B^{(i,i+1)}_{4}B^{(i+1,i+2)}_{1}+
B^{(i,i+1)}_{2}B^{(i+1,i+2)}_{3}+B^{(i,i+1)}_{3}B^{(i+1,i+2)}_{2},
\end{eqnarray}
\end{widetext}
independently on the off-diagonal elements.
If the swapping procedure is carried out many times with an ensemble of
pairs distributed along the link, 
the iterative rule (\ref{swapp}) describes the evolution of fidelities in the
link with the switchers.
However, the entanglement swapping of
weakly entangled states leads to generation of lower entangled
pair at a large distance.
To improve this long-distance entanglement,
we can use LOCC purification procedure
\cite{Deutsch} requiring multiple copies of the long-distance states.
Deutsch's purification allows us to distill a
state with needed amount of entanglement from $M$ imperfectly entangled pairs
$\rho^{(0,L+1)}$ having fidelity $B_{1}>1/2$.  
It is composed from three steps: (i) the local unitary
operations on the sender side
\begin{equation}
|0\rangle \rightarrow  \frac{1}{\sqrt{2}}(|0\rangle-i|1\rangle),
\hspace{0.3cm}
|1\rangle \rightarrow  \frac{1}{\sqrt{2}}(|1\rangle-i|0\rangle),
\end{equation}
and on the receiver side
\begin{equation}
|0\rangle \rightarrow  \frac{1}{\sqrt{2}}(|0\rangle+i|1\rangle),
\hspace{0.3cm}
|1\rangle \rightarrow  \frac{1}{\sqrt{2}}(|1\rangle+i|0\rangle),
\end{equation}
followed by (ii) action of the quantum C-NOT operation
on the both sides
\begin{equation}
|a\rangle_{C}|b\rangle_{T}\rightarrow |a\rangle_{C}|a\oplus b\rangle_{T}
\end{equation}
where $C$ is control qubit and $T$ is target qubit, and (iii) measurement of
the target qubits on the both sides. If the measurement
outcomes coincide, the control
pair is kept for next round. For two different states
$\rho^{(0,L+1)}$ and $\rho'^{(0,L+1)}$ described by fidelities $B_{i}$
and $B'_{i}$, the distilled pair $\bar{\rho}^{(0,l+1)}$
has the following shifted fidelities
\begin{eqnarray}\label{deutsch}
{\bar B_{1}}&=&\frac{1}{N}(B_{1}B'_{1}+B_{4}'B_{4}),\hspace{0.2cm}
{\bar B_{2}}=\frac{1}{N}(B_{1}B'_{4}+B_{4}'B_{1}),\nonumber\\
{\bar B_{3}}&=&\frac{1}{N}(B_{2}B'_{2}+B_{3}'B_{3}),\hspace{0.2cm}
{\bar B_{4}}=\frac{1}{N}(B_{2}B'_{3}+B_{2}'B_{3}),\nonumber\\
\end{eqnarray}
where $N=(B_{1}+B_{4})(B'_{1}+B'_{4})+(B_{2}+B_{3})(B'_{2}+B'_{3})$, 
$B_{i},B'_{i}$ are the fidelities of input pairs and $\bar{B}_{i}$ are
the fidelities of an output distilled pair.
Particularly, if $B_{1}>1/2$ for all shared states, there is only 
one fixed point for $B_{1}=1$ \cite{Macchiavello98}. 
After some steps of procedure, the outgoing state is
sufficiently close to this fixed point (as it is necessary) and the sender
and receiver establish a resource for key distribution with a given security.
After $m$ repetition of purification procedure, we can
generate a strongly entangled pair from $2^{m}$ input lower entangled pairs.
In this way, an eavesdropping
attack is {\em factorized} from the sender-receiver state.
On the other hand, if the purification procedure does not converge, we 
cannot use the outcoming pairs for the key distribution. Thus, a
convergence of the procedure must be simultaneously proved in 
the communication.
Note, when there are the imperfect local operations, only a maximal
non-unit fidelity is obtained, however as has been recently proved,
the users may nevertheless use
these pairs for secure communication \cite{Aschauer02}.


The iterative procedures (\ref{swapp}) and (\ref{deutsch})
represent a complete solution of the presented problem for any
two-qubit states. To discuss the security costs in dependence on the number
of switchers, we will analyse two interesting cases of the decoherence
process. As first, we will analyse the security costs in the network
with the states $\bar{\rho}^{(i,i+1)}$ after
an optimal individual eavesdropping attack -
quantum nondemolition (QND) measurement.
We assume that state $|B_{1}\rangle$ is decohered by the standard
QND monitoring of basis states $|0\rangle$ and $|1\rangle$
\begin{eqnarray}
|00\rangle|0\rangle_{E} &\rightarrow &
|00\rangle|0\rangle_{E},\nonumber\\
|11\rangle|0\rangle_{E} &\rightarrow &
|11\rangle (R|0\rangle_{E}+\sqrt{1-R^{2}}|1\rangle_{E}),
\end{eqnarray}
where $R$ is a measure of robustness against decoherence and
$|0\rangle_{E}$ and $|1\rangle_{E}$ are basis states of the environment.
After monitoring process and tracing out the environmental states,
we obtain input states for swapping procedure
\begin{equation}\label{qndst}
\rho^{(i,i+1)}=B_{1}|B_{1}\rangle\langle B_{1}|+(1-B_{1})|B_{2}\rangle\langle
B_{2}|
\end{equation}
where $B_{1}=\frac{1}{2}(1+R)$ is fidelity determined by the robustness.
Note, that state (\ref{qndst}) is entangled and nonlocal for
every $R>0$. During swapping procedure (\ref{swapp}),
the structure of state (\ref{qndst}) is preserved
and only the fidelity $B_{1}$ changes according to
\begin{eqnarray}\label{sw2}
B^{(i,i+2)}_{1}&=&B^{(i,i+1)}_{1}B^{(i+1,i+2)}_{1}+\nonumber\\
& &+(1-B^{(i,i+1)}_{1})(1-B^{(i+1,i+2)}_{1}).
\end{eqnarray}
This rule can be iterated to obtain an outgoing state from swapping procedure.
Employing robustness parameter $R$, we can simply find that outgoing
robustness is $R'=R^{L}$, which is signature of pronounced
{\em exponential} decoherence.
After swapping procedure (\ref{sw2}),
the resulting state is considered as the input of Deutsch's
purification procedure for $F=B_{1}$
\begin{equation}
\bar{F}=\frac{FF'}{FF'+(1-F)(1-F')}
\end{equation}
which can be simplified for the same states utilizing the robustness
$\bar{R}=2R/(1+R^{2})$.
Introducing two new parameters $r$ and $r'$,
$R=\tanh r$ and $\bar{R}=\tanh\bar{r}$,
we have $\bar{r}=2^{m}r=Mr$, after $m$ levels of the
purification procedure.
Due to this relation, we can express a relation between 
near-users robustness $R$ and the long-distance robustness 
in form of $\bar{R}=\tanh(M\mbox{arctanh} R^{L})$ and approximately
determine a lower bound on $M$ as
\begin{equation}
M\approx\mbox{Int}\left(\frac{\ln (1+R)-\ln (1-R)}{\ln (1+R^{L})-
\ln (1-R^{L})}\right)+1.
\end{equation}
If $R>0$, then purification procedure converges to fixed point
$\bar{B}_{1}=1$ for every $L$,
however the number of needed entangled
states {\em exponentially} increases, as is depicted in Fig.~2 for
almost maximal ``working'' fidelities $B_{1}=0.9925,0.985,0.9625$.
This exponential overhead is necessary cost that must be paid to
establish a security of quantum channel at a large distance. It is still opened
question, whether the LOCC purification
procedure can be modified in such a way to obtain only a
sub-exponential overhead.


As a second example, we consider Werner state between the nearest
users. Note, that Werner state results
from the simplest eavesdropping strategy on
the state $|B_{1}\rangle$:
Eve steals the photon transmitted from sender station with
probability $p$ and subsequently, sends another photon with 
random polarization towards receiver station. 
Particularly, for the $L$ switchers with the $L+1$ Werner states 
(having $B_{2}=B_{3}=B_{4}$)
\begin{equation}\label{Werner}
\rho^{(i,j)}=p|B_{1}\rangle\langle B_{1}|+
\frac{1-p}{4}1\otimes 1,
\end{equation}
where $p=(4B_{1}-1)/3$, an outgoing state is the Werner state
(\ref{Werner}) with $p'=p^{L}$.
This is a signature a {\em exponential}
decrease of nonlocality and entanglement in a network with large $L$,
even if particular $p$ approaches unity.
Note, that Werner state (\ref{Werner}) is entangled if the entanglement
factor $\Lambda=1/2(1-3p)$ is negative and does not admit a 
local realistic explanation if the 
Bell factor $B_{\rm max}=2\sqrt{2}p$ is larger than $2$.
After $L$ swapping procedures, we obtain a long distance Werner state
with the following fidelity
\begin{equation}
B'_{1}=\frac{3}{4}\left(\frac{4}{3}B_{1}-\frac{1}{3}\right)^{L}+
\frac{1}{4}.
\end{equation}
which is entangled if and only if
$B'_{1}>\frac{1}{2}$. Then if is possible to use the purification
procedure, i.e. numerically iterate the map (\ref{deutsch}).
Number $M=2^{m}$ (where $m$ is the number of purification steps)
of pairs needed to achieve the same
fidelity, as has been
between the nearest users,
is depicted in Fig.~3 in dependence on the number $L$ of the
switchers for the ``working'' fidelities
$B_{1}=0.9925,0.985,0.9625$.
As can be seen, we have two distinct regions of overhead:
for given $B_{1}$,
the number $M$ firstly {\em exponentially} grows with $L$ and 
this divergence changes to rapid {\em super-exponential} behavior.
Note, that super-exponential
overhead is pronounced as the fidelity $B_{1}$ approaches unity.
For sufficiently large $B_{1}>0.95$, 
we can optimize NPP to prevent super-exponential overhead, 
if we restrict the number $L$ of switchers between two repeaters
\begin{equation}\label{restr}
L<\frac{1}{2\left(1-\frac{\ln (4B_{1}-1)}{\ln 3}\right)}, 
\end{equation}
to half of the maximal value $L_{max}$,
where $L_{max}$ is depicted in Fig.~2 by a cross symbol.
In addition, it can be numerically checked that the analysed case,
employing the Werner states, is the most expensive from all the cases with
the same state (\ref{genst}) between
the nearest users.
Because we are not able to priori know
what are the states distributed between nearest users,
the restriction (\ref{restr})
is a general way how to prevent super-exponential overhead and 
it represents {\em optimized} nested purification protocol (ONPP).

In summary, irrespective to very encouraging
experimental results in entangled state cryptography \cite{Naik00},
entanglement swapping procedure \cite{swapp},
a strong {\em exponential} decoherence
arises at a large distance
from ``multiplication'' of small deviations from perfect
near-users entangled states. To avoid this decoherence, we can use the
nested purification procedure \cite{Briegel98,Dur99}.
In this paper, we have found a general formula for
the fidelity evolution in this
nested purification procedure and
we have demonstrated that, per nesting segment, this procedure
exhibits {\em exponential} or even {\em super-exponential}
overhead in the entanglement resources, 
in dependence on the number of the entanglement switchers.
We analysed the case with a maximal overhead
and optimized NPP by a restriction
of maximal number of switchers between two repeaters 
to avoid {\em super-exponential} overhead.
It is important for any construction of the sufficiently
secure large cryptographic networks in the next future.
However, there is a still opened question
whether it is possible to obtain a sub-exponential overhead
by a modification of swapping/purification procedure.

I would like to thank J. Fiur\' a\v sek for interesting discussions
about this problem.
This research was supported under the project LN00A015 of the Ministry
of Education of the Czech Republic.


\newpage

\begin{figure}\label{ent0}
\vspace{1cm}
\caption{Segment of nested purification protocol: R- quantum repeaters,
 S- switchers based on entanglement swapping employed by particular users.}
\end{figure}

\begin{figure}\label{ent2}
\caption{Number of entangled pairs $M=2^{m}$ needed for distillation of
one entangled pair between distant users
with the same security as for the nearest users, $L$
number of the switchers in the link mutually sharing the same QND
states with given $R$: (a) R=0.985 ($B_{1}=0.9925$), (b)
R=0.97 ($B_{1}=0.9850$), (c) R=0.925 ($B_{1}=0.9625$).}
\end{figure}

\begin{figure}\label{ent1}
\caption{Number of entangled pairs $M=2^{m}$ needed for distillation of
one entangled pair between distant users
with the same security as for the nearest users, $L$
number of the switchers in the link muttually sharing the same Werner
states with given $p$: (a) p=0.99 ($B_{1}=0.9925$), (b)
p=0.98 ($B_{1}=0.9850$), (c) p=0.95 ($B_{1}=0.9625$). The crosses denote 
a threshold number $L_{max}$ of the switchers between two repeaters.}
\end{figure}

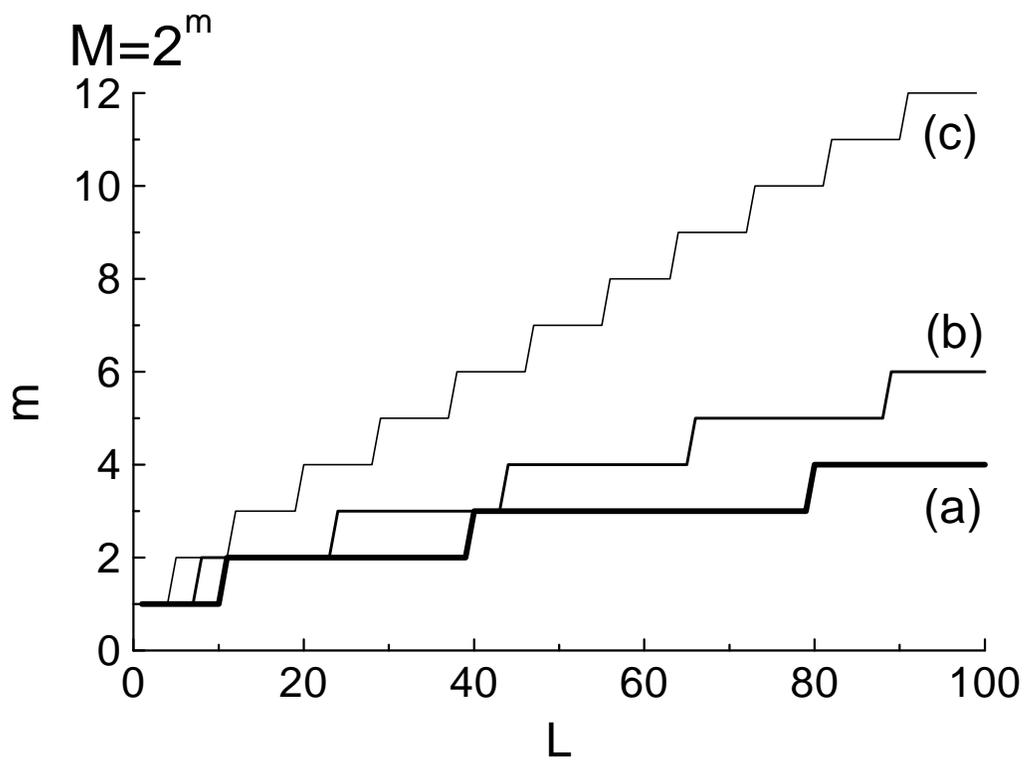
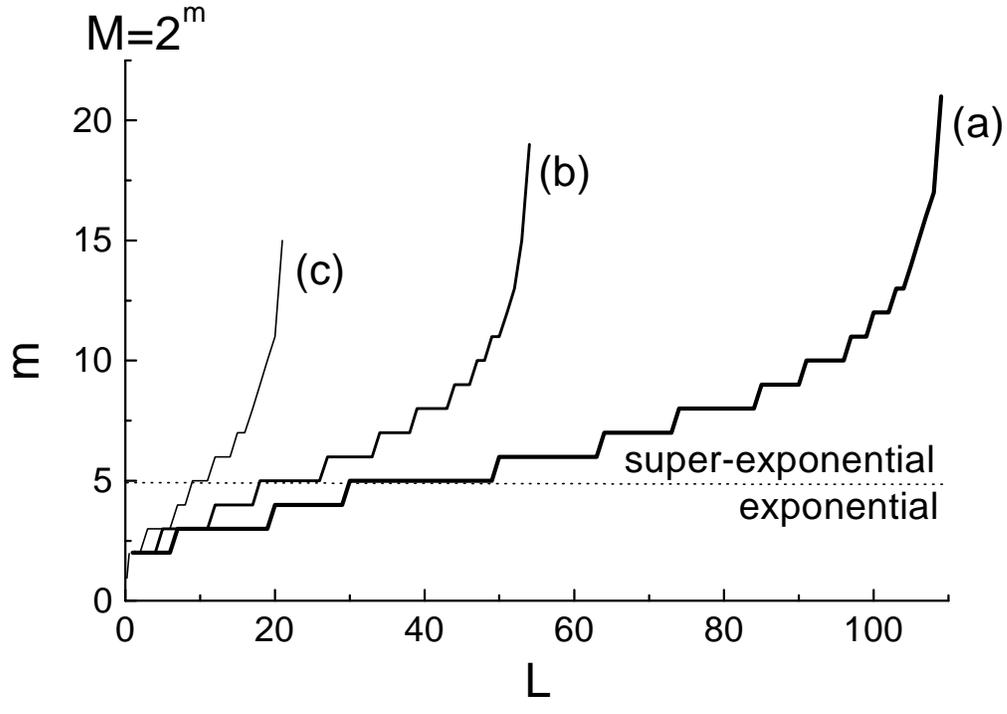
\end{document}